\documentclass[12pt]{article}

\renewcommand{\baselinestretch}{1.5}

\newcommand{\bl}{\mbox{\boldmath$l$}}
\newcommand{\pmbpi}{\,^{\pm}\!\mbox{\boldmath$\pi$}}
\renewcommand{\d}{{\rm d}}
\newcommand{\ppi}{\,^+\!\pi}
\newcommand{\mpi}{\,^-\!\pi}
\newcommand{\pmu}{\,^+\!\mu}
\newcommand{\mmu}{\,^-\!\mu}
\newcommand{\pSig}{\,^+\!\Sigma}
\newcommand{\mSig}{\,^-\!\Sigma}
\newcommand{\pmSig}{\,^{\pm}\!\Sigma}
\newcommand{\R}{\bar{R}}
\newcommand{\Om}{\bar{\Omega}}
\newcommand{\pmR}{\,^{\pm}\!R}
\newcommand{\Rsig}{R_{\sigma^2}}
\newcommand{\pmRsig}{\,^{\pm}\!R_{\sigma^2}}
\newcommand{\pmcalN}{\,^{\pm}\!{\cal N}}
\newcommand{\pcalN}{\,^{+}\!{\cal N}}
\newcommand{\mcalN}{\,^{-}\!{\cal N}}
\newcommand{\pmm}{\,^{\pm}\!m}
\newcommand{\pmbarm}{\,^{\pm}\!\bar{m}}
\newcommand{\pmpi}{\,^{\pm}\!\pi}
\newcommand{\pmmu}{\,^{\pm}\!\mu}
\newcommand{\pmbphi}{\,^{\pm}\!\mbox{\boldmath$\phi$}}
\newcommand{\pmbSig}{\,^{\pm}\!\mbox{\boldmath$\Sigma$}}
\newcommand{\pmphi}{\,^{\pm}\!\phi}
\newcommand{\pmeta}{\,^{\pm}\!\eta}
\newcommand{\pmtheta}{\,^{\pm}\!\theta}
\newcommand{\pmtau}{\,^{\pm}\!\tau}
\newcommand{\pmzeta}{\,^{\pm}\!\zeta}
\newcommand{\pmalpha}{\,^{\pm}\!\alpha}
\newcommand{\pmchi}{\,^{\pm}\!\chi}
\newcommand{\pmbn}{\,^{\pm}\!\mbox{\boldmath$n$}}
\newcommand{\pmbnsig}{\,^{\pm}\!\mbox{\boldmath$n$}_{\sigma^2}}
\newcommand{\pmbeonesig}{\,^{\pm}\!\mbox{\boldmath$e$}_{\!1\sigma^2}}
\newcommand{\pmbetwosig}{\,^{\pm}\!\mbox{\boldmath$e$}_{2\sigma^2}}
\newcommand{\pmbm}{\,^{\pm}\!\mbox{\boldmath$m$}}
\newcommand{\pmmsig}{\,^{\pm}\!m_{\sigma^2}}
\newcommand{\pvsig}{\,^{+}\!v_{\sigma^2}}
\newcommand{\mvsig}{\,^{-}\!v_{\sigma^2}}
\newcommand{\pmbarmsig}{\,^{\pm}\!\bar{m}_{\sigma^2}}
\newcommand{\tausig}{\tau_{\sigma^2}}
\newcommand{\msig}{m_{\sigma^2}}
\newcommand{\pisig}{\pi_{\sigma^2}}
\newcommand{\pmzetasig}{\,^{\pm}\!\zeta_{\sigma^2}}
\newcommand{\pmetasig}{\,^{\pm}\!\eta_{\sigma^2}}
\newcommand{\pmchisig}{\,^{\pm}\!\chi_{\sigma^2}}
\newcommand{\pmbpisig}{\,^{\pm}\!\mbox{\boldmath$\pi$}_{\sigma^2}}
\newcommand{\sq}{{\cal S}q}
\newcommand{\T}{\bar{T}}
\newcommand{\U}{\bar{U}}

\begin{document}

\title{On positivity of quantum measure and of effective action in area tensor
 Regge calculus}
\author{V.M. Khatsymovsky \\
 {\em Budker Institute of Nuclear Physics} \\ {\em
 Novosibirsk,
 630090,
 Russia}
\\ {\em E-mail address: khatsym@inp.nsk.su}}
\date{}
\maketitle
\begin{abstract}
Because of unboundedness of the general relativity action, Euclidean version of the
path integral in general relativity requires definition. Area tensor Regge calculus is
considered in the representation with independent area tensor and finite rotation
matrices. Being integrated over rotation matrices the path integral measure in area
tensor Regge calculus is rewritten by moving integration contours to complex plain so
that it looks as that one with effective action in the exponential with positive real
part. We speculate that positivity of the measure can be expected in the most part of
range of variation of area tensors.
\end{abstract}


\newpage

The formal nonrenormalisability of quantum version of general relativity (GR) may
cause us to try to find alternatives to the continuum description of underlying
spacetime structure. An example of such the alternative description may be given by
Regge calculus (RC) suggested in 1961 \cite{Regge}. It is the exact GR developed in
the piecewise flat spacetime which is a particular case of general Riemannian
spacetime \cite{Fried}. In turn, the general Riemannian spacetime can be considered as
limiting case of the piecewise flat spacetime \cite{Fein}. Any piecewise flat
spacetime is simlicial one: it can be represented as collection of a (countable)
number of the flat 4-dimensional {\it simplices}(tetrahedrons), and its geometry is
completely specified by the countable number of the freely chosen lengths of all edges
(or 1-simplices). Thus, RC implies a {\it discrete} description {\it alternative} to
the usual continuum one. For a review of RC and alternative discrete gravity
approaches see, e. g., \cite{RegWil}.

The discrete nature of the simplicial description presents a difficulty in the
(canonical) quantization of such the theory due to the absence of a regular continuous
coordinate playing the role of time. Therefore one cannot immediately develop
Hamiltonian formalism and canonical (Dirac) quantization. To do this we need to return
to the partially continuum description, namely, with respect to only one direction
shrinking sizes of all the simplices along this direction to those infinitely close to
zero. The linklengths and other geometrical quantities become functions of the
continuous coordinate taken along this direction. We can call this coordinate time $t$
and develop quantization procedure with respect to this time. The result of this
procedure can be formulated as some path integral measure. It is quite natural to
consider this measure as a (appropriately defined) limiting continuous time form of a
measure on the set of the original completely discrete simplicial spacetimes. This
last completely discrete measure is just the object of interest to be found. The
requirement for this measure to have the known limiting continuous time form can be
considered as a starting postulate in our construction. The issuing principles are of
course not unique, and another approaches to defining quantum measure in RC based on
another physical principles do exist \cite{HamWil1,HamWil2}.

The above condition for the completely discrete measure to possess required continuous
time limit does not defines it uniquely as long as only one fixed direction which
defines $t$ is considered. However, different coordinate directions should be
equivalent and we have a right to require for the measure to result in the canonical
quantization measure in the continuous time limit {\it whatever} coordinate direction
is chosen to define a time. These requirements are on the contrary a priori too
stringent, and it is important that on some configuration superspace (extended in
comparison with superspace of the genuine simplicial geometries) such the measure
turns out to exist.

Briefly speaking, we should, first, find continuous time limit for Regge action,
recast it in the canonical Hamiltonian form and write out the Hamiltonian path
integral, the measure in the latter being called for a moment the continuous time
measure; second, we should check for existence and (if exists) find the measure
obeying the property to tend in the continuous time limit (with concept "to tend"
being properly defined) to the found continuous time measure irrespectively of the
choice of the time coordinate direction. When passing to the continuous time RC we are
faced with the difficulty that the description of the infinitely flattened in some
direction simplex purely in terms of the lengths is singular.

The way to avoid singularities in the continuous time limit is to extend the set of
variables via adding the new ones having the sense of angles and considered as
independent variables. Such the variables are the finite rotation matrices which are
the discrete analogs of the connections in the continuum GR. The situation considered
is analogous to that one occurred when recasting the Einstein action in the
Hilbert-Palatini form,
\begin{equation}                                                                    
\label{S-HilPal} {1\over 2}\int{R\sqrt{g}{\rm d}^4x} \Leftarrow {1\over
8}\int{\epsilon_{abcd}\epsilon^{\lambda\mu\nu\rho}e^a_{\lambda}e^b_{\mu}
[\partial_{\nu}+\omega_{\nu},\partial_{\rho}+\omega_{\rho}]^{cd}{\rm d}^4x},
\end{equation}

\noindent where the tetrad $e^a_{\lambda}$ and connection $\omega^{ab}_{\lambda}$ =
$-\omega^{ba}_{\lambda}$ are independent variables, the RHS being reduced to LHS in
terms of $g_{\lambda\mu}$ = $e^a_{\lambda}e_{a\mu}$ if we substitute for
$\omega^{ab}_{\lambda}$ solution of the equations of motion for these variables in
terms of $e^a_{\lambda}$. The Latin indices $a$, $b$, $c$, ... are the vector ones
with respect to the local Euclidean frames which are introduced at each point $x$.

Now in RC the Einstein action in the LHS of (\ref{S-HilPal}) becomes the Regge action,
\begin{equation}                                                                    
\label{S-Regge} \sum_{\sigma^2}{\alpha_{\sigma^2}|\sigma^2|},
\end{equation}

\noindent where $|\sigma^2|$ is the area of a triangle (the 2-simplex) $\sigma^2$,
$\alpha_{\sigma^2}$ is the angle defect on this triangle, and summation run over all
the 2-simplices $\sigma^2$. The discrete analogs of the tetrad and connection, edge
vectors and finite rotation matrices, were first considered in \cite{Fro}. The local
Euclidean frames live in the 4-simplices now, and the analogs of the connection are
defined on the 3-simplices $\sigma^3$ and are the matrices $\Omega_{\sigma^3}$
connecting the frames of the pairs of the 4-simplices $\sigma^4$ sharing the 3-faces
$\sigma^3$. These matrices are the finite SO(4) rotations in the Euclidean case (or
SO(3,1) rotations in the Lorentzian case) in contrast with the continuum connections
$\omega^{ab}_{\lambda}$ which are the elements of the Lee algebra so(4)(so(3,1)) of
this group. This definition includes pointing out the direction in which the
connection $\Omega_{\sigma^3}$ acts (and, correspondingly, the opposite direction, in
which the $\Omega^{-1}_{\sigma^3}$ = $\bar{\Omega}_{\sigma^3}$ acts), that is, the
connections $\Omega$ are defined on the {\it oriented} 3-simplices $\sigma^3$. Instead
of RHS of (\ref{S-HilPal}) we use exact representation which we suggest in our work
\cite{Kha1},
\begin{equation}                                                                    
\label{S-RegCon}%
S(v,\Omega) = \sum_{\sigma^2}{|v_{\sigma^2}|\arcsin{v_{\sigma^2}\circ
R_{\sigma^2}(\Omega)\over |v_{\sigma^2}|}}
\end{equation}

\noindent where we have defined $A\circ B$ = ${1\over 2}A^{ab}B_{ab}$, $|A|$ =
$(A\circ A)^{1/2}$ for the two tensors $A$, $B$; $v_{\sigma^2}$ is the dual bivector
of the triangle $\sigma^2$ in terms of the vectors of its edges $l^a_1$, $l^a_2$,
\begin{equation}                                                                    
\label{v=ll} v_{\sigma^2ab} = {1\over 2}\epsilon_{abcd}l^c_1l^d_2
\end{equation}

\noindent (in some 4-simplex frame containing $\sigma^2$). The curvature matrix
$R_{\sigma^2}$ on the 2-simplex $\sigma^2$ is the path ordered product of the
connections $\Omega^{\pm 1}_{\sigma^3}$ on the 3-simplices $\sigma^3$ sharing
$\sigma^2$ along the contour enclosing $\sigma^2$ once and contained in the
4-simplices sharing $\sigma^2$,
\begin{equation}                                                                    
\label{R-Omega} R_{\sigma^2} = \prod_{\sigma^3\supset\sigma^2}{\Omega^{\pm
1}_{\sigma^3}}.
\end{equation}

\noindent  As we can show, when substituting as $\Omega_{\sigma^3}$ the genuine
rotations connecting the neighbouring local frames as functions of the genuine Regge
lengths into the equations of motion for $\Omega_{\sigma^3}$ with the action
(\ref{S-RegCon}) we get exactly the closure condition for the surface of the 3-simplex
$\sigma^3$ (vanishing the sum of the bivectors of its 2-faces) written in the frame of
one of the 4-simplices containing $\sigma^3$, that is, the identity. This means that
(\ref{S-RegCon}) is the exact representation for (\ref{S-Regge}). At the same time,
general solution to the equations of motion is wider than that leading to
$R_{\sigma^2}(\Omega)$ rotating around $\sigma^2$ by the defect angle
$\alpha_{\sigma^2}$.

We can pass to the continuous time limit in (\ref{S-RegCon}) in a nonsingular manner
and recast it to the canonical (Hamiltonian) form \cite{Kha2}. This allows us to write
out Hamiltonian path integral. The above problem of finding the measure which results
in the Hamiltonian path integral measure in the continuous time limit whatever
coordinate is chosen as time has solution in 3 dimensions \cite{Kha3}. A specific
feature of the 3D case important for that is commutativity of the dynamical
constraints leading to a simple form of the functional integral. The 3D action looks
like (\ref{S-RegCon}) with area tensors $v_{\sigma^2}$ substituted by the egde vectors
$\bl_{\sigma^1}$ independent of each other. In 4 dimensions, the variables
$v_{\sigma^2}$ are not independent but obey a set of (bilinear) {\it intersection
relations}. For example, tensors of the two triangles $\sigma^2_1$, $\sigma^2_2$
sharing an edge satisfy the relation
\begin{equation}                                                                    
\label{v*v}%
\epsilon_{abcd}v^{ab}_{\sigma^2_1}v^{cd}_{\sigma^2_2} = 0.
\end{equation}

\noindent These purely geometrical relations can be called kinematical constraints.
The idea is to construct quantum measure first for the system with formally
independent area tensors. That is, originally we concentrate on quantization of the
dynamics while kinematical relations of the type (\ref{v*v}) are taken into account at
the second stage. Note that the RC with formally independent (scalar) areas have been
considered in the literature \cite{RegWil,BarRocWil}.

The theory with formally independent area tensors can be called area tensor RC.
Consider the Euclidean case. The Einstein action is not bounded from below, therefore
the Euclidean path integral itself requires careful definition. Our result for the
constructed in the above way completely discrete quantum measure \cite{Kha4} can be
written as a result for vacuum expectations of the functions of the field variables
$v$, $\Omega$. Upon passing to integration over imaginary areas with the help of the
formal replacement of the tensors of a certain subset of areas $\pi$ over which
integration in the path integral is to be performed, $$\pi \rightarrow -i\pi,$$ the
result reads
\begin{eqnarray}                                                                    
\label{VEV2}%
<\Psi (\{\pi\},\{\Omega\})> & = & \int{\Psi (-i\{\pi\}, \{\Omega\})\exp{\left (-\!
\sum_{\stackrel{t-{\rm like}}{\sigma^2}}{\tau _{\sigma^2}\circ
R_{\sigma^2}(\Omega)}\right )}}\nonumber\\
 & & \hspace{-20mm} \cdot \exp{\left (i
\!\sum_{\stackrel{\stackrel{\rm not}{t-{\rm like}}}{\sigma^2}} {\pi_{\sigma^2}\circ
R_{\sigma^2}(\Omega)}\right )}\prod_{\stackrel{\stackrel{\rm
 not}{t-{\rm like}}}{\sigma^2}}{\rm d}^6
\pi_{\sigma^2}\prod_{\sigma^3}{{\cal D}\Omega_{\sigma^3}} \nonumber\\ & \equiv &
\int{\Psi (-i\{\pi\},\{\Omega\}){\rm d} \mu_{\rm area}(-i\{\pi\},\{\Omega\})},
\end{eqnarray}

\noindent where ${\cal D}\Omega_{\sigma^3}$ is the Haar measure on the group SO(4) of
connection matrices $\Omega_{\sigma^3}$. Appearance of some set ${\cal F}$ of
triangles $\sigma^2$ integration over area tensors of which is omitted (denoted as
"$t$-like" in (\ref{VEV2}))is connected with that integration over {\it all} area
tensors is generally infinite, in particular, when normalizing measure (finding
$<1>$). Indeed, different $R_{\sigma^2}$ for $\sigma^2$ meeting at a given link
$\sigma^1$ are connected by Bianchi identities \cite{Regge}. Therefore for the
spacetime of Minkowsky signature (when exponent is oscillating over all the area
tensors) the product of $\delta^6(R_{\sigma^2} - \R_{\sigma^2})$ for all these
$\sigma^2$ which follow upon integration over area tensors for these $\sigma^2$
contains singularity of the type of $\delta$-function squared. To avoid this
singularity we should confine ourselves by only integration over area tensors on those
$\sigma^2$ on which $R_{\sigma^2}$ are independent, and complement ${\cal F}$ to this
set of $\sigma^2$ are those $\sigma^2$ on which $R_{\sigma^2}$ are by means of the
Bianchi identities functions of these independent $R_{\sigma^2}$. Let us adopt regular
way of constructing 4D simplicial structure of the 3D simplicial geometries (leaves)
of the same structure. Denote by $A$, $B$, $C$, ... vertices of the 4D simplicial
complex while $n$-simplex $\sigma^n$ is denoted by the set of its $n + 1$ vertices in
round brackets (unordered sequence), $(A_1A_2...)$. The $i$, $k$, $l$, ... are
vertices of the current leaf, $i^+$, $k^+$, $l^+$, ... and $i^-$, $k^-$, $l^-$, ...
are corresponding vertices of the nearest future and past in $t$ leaves. Or, dealing
with Euclidean time, we shall speak of the "upper" and "lower" leaves, respectively.
Each vertex is connected by links (edges) with its $\pm$-images. These links (of the
type of $(ii^+)$, $(ii^-)$) will be called $t$-{\it like} ones (do not mix with the
term "timelike" which is reserved for the local frame components). The {\it leaf}
links $(ik)$ are completely contained in the 3D leaf. There may be {\it diagonal}
links $(ik^+)$, $(ik^-)$ connecting a vertex with the $\pm$-images of its neighbors.
We call arbitrary simplex $t$-{\it like} one if it has $t$-like edge, the {\it leaf}
one if it is completely contained in the 3D leaf and {\it diagonal} one in other
cases. It can be seen that the set of the $t$-like triangles is fit for the role of
the above set ${\cal F}$. In the case of general 4D simplicial structure we can deduce
that the set ${\cal F}$ of the triangles with the Bianchi-dependent curvatures pick
out some one-dimensional field of links, and we can simply take it as definition of
the coordinate $t$ direction so that ${\cal F}$ be just the set of the $t$-like
triangles. Also existence of the set ${\cal F}$ naturally fits our initial requirement
that limiting form of the full discrete measure when any one of the coordinates (not
necessarily $t$!) is made continuous by flattening the 4-simplices in the
corresponding direction should coincide with Hamiltonian path integral (with that
coordinate playing the role of time). Namely, in the Hamiltonian formalism absence of
integration over area tensors of triangles which pick out some coordinate $t$
($t$-like ones) corresponds to some gauge fixing.

There is the invariant (Haar) measure ${\cal D}\Omega$ in (\ref{VEV2}) which looks
natural from symmetry considerations. From the formal point of view, in the
Hamiltonian formalism (when one of the coordinates is made continuous) this arises
when we write out standard Hamiltonian path integral for the Lagrangian with the
kinetic term $\pi_{\sigma^2}\circ\Om_{\sigma^2}\dot{\Omega}_{\sigma^2}$
\cite{Kha3,Kha4}. To this end, one might pass to the variables $\Omega_{\sigma^2}
\pi_{\sigma^2}\!$ = $\!P_{\sigma^2}$ and $\Omega_{\sigma^2}$ (in 3D case used in
\cite{Wael,Kha3}). The kinetic term $P\dot{\Omega}$ with arbitrary matrices $P$,
$\Omega$ leads to the standard measure ${\rm d}^{16}P{\rm d}^{16}\Omega$, but there
are also $\delta$-functions taking into account II class constraints to which $P$,
$\Omega$ are subject, $\delta^{10}(\Om\Omega - 1)\!$ $\!\delta^{10}(\Om P +
\bar{P}\Omega)$. Integrating out these just gives ${\rm d}^6\pi{\cal D}^6\Omega$.
Following our strategy of recovering full discrete measure from requirement that it
reduces to the Hamiltonian path integral whatever coordinate is made continuous, the
same Haar measure should be present also in the full discrete measure.

One else specific feature of the quantum measure is the absence of the inverse
trigonometric function 'arcsin' in the exponential, whereas the Regge action
($\ref{S-RegCon}$) contains such functions. This is connected with using the canonical
quantization at the intermediate stage of derivation: in gravity this quantization is
completely defined by the constraints, the latter being equivalent to those ones
without $\arcsin$ (in some sense on-shell).

In what follows, it is convenient to split antisymmetric matrices ($\pi$ and generator
of $R$) into self- and antiselfdual parts, then the measure (\ref{VEV2}) splits into
two factors, in the self- and antiselfdual sectors,
\begin{eqnarray}                                                                    
\pi_{ab} & \equiv & {1\over 2}\ppi_k\pSig^k_{ab} + {1\over 2}\mpi_k\mSig^k_{ab} \\%
\pmR & = & \exp (\pmbphi\pmbSig) = \cos \!\pmphi + \pmbSig\pmbn\sin \!\pmphi \\     
\d \mu_{\rm area} & = & \d \pmu_{\rm area} \d \mmu_{\rm area}.                     
\end{eqnarray}

\noindent Here $\pmbn$ = $\pmbphi / \pmphi$ is unit vector and the basis of self- and
antiselfdual matrices $i\pmSig^k_{ab}$ obeys the Pauli matrix algebra.

Since as pointed out below the eqs. (\ref{S-RegCon}) - (\ref{R-Omega}) the classical
equations of motion for $\Omega$ do not restrict the resulting $R_{\sigma^2}(\Omega )$
be exactly the rotation around $\sigma^2$ by the defect angle $\alpha_{\sigma^2}$, the
sense of $\Omega$, $R(\Omega)$ as physical observables is restricted. Consider
averaging functions of only area tensors $\pi_{\sigma^2}$. By the properties of
invariant measure, integrations over $\prod{\cal D}\Omega_{\sigma^3}$ in (\ref{VEV2})
reduce to integrations over $\prod{\cal D}R_{\sigma^2}$ with independent
$R_{\sigma^2}$ (i. e. $\sigma^2$ are just not $t$-like) and some number of connections
$\prod{\cal D}\Omega_{\sigma^3}$ which we can call gauge ones. The expectation value
of any field monomial, $<\pi^{a_1b_1}_{\sigma^2_1}...\pi^{a_nb_n}_{\sigma^2_n}>$
reduces to the (derivatives of) $\delta$-functions $\delta (R^{a_ib_i}_{\sigma^2_i} -
R^{b_ia_i}_{\sigma^2_i} )$ which are then integrated out over ${\cal D}R_{\sigma^2_i}$
giving finite nonzero answer. This is consequence of i) the underlying Dirac-Hamilton
principle of quantization (leading to $\d^6\pi_{\sigma^2}{\cal D}R_{\sigma^2}$ in the
measure) and of ii) conception of independent area tensors (integrations over
$\d^6\pi_{\sigma^2}$ are independent leading to $\delta$-functions). This holds in the
Minkowsky spacetime as well (and in the first instance since oscillating exponent is
present there from the very beginning). The Euclidean expectations values correspond
to the Minkowskian ones in the spacelike region. The formal passing to the Euclidean
version by simply writing $\exp (-\pi_{\sigma^2}\circ R_{\sigma^2})$ in the measure
(not with additional substitution $\pi_{\sigma^2}\!$ $\rightarrow$
$\!-i\pi_{\sigma^2}$ in the integration variables as in (\ref{VEV2})) might result,
upon integrating over ${\cal D}R_{\sigma^2}$, in appearance of the terms with both
factors, $\exp (+|\pmbpi_{\sigma^2}|)$ and $\exp (-|\pmbpi_{\sigma^2}|)$. This is
consequence of that iii) $R_{\sigma^2}$ are {\it finite} SO(4) rotations, not elements
of Lee group so(4) - therefore the stationary phase points in the integrals  over
${\cal D}R_{\sigma^2_i}$ correspond just to $\pmpi_{\sigma^2}\circ\pmR_{\sigma^2}\!$ =
$\!+|\pmbpi_{\sigma^2}|$ and $\pmpi_{\sigma^2}\circ\pmR_{\sigma^2}\!$ =
$\!-|\pmbpi_{\sigma^2}|$. Due to the above mentioned finiteness of area monomial VEVs
the growing exponents $\exp (+|\pmbpi_{\sigma^2}|)$ should be excluded. Thus, the
measure upon integration over connections should exponentially decrease with areas as
$\exp (-|\pmbpi_{\sigma^2}|)$. Once again, collect the reasons for that,\\%
i) Dirac-Hamiltonian canonical quantization;\\%
ii) conception of independent area tensors;\\%
iii) connection matrices being finite SO(4) rotations, not elements of the Lee group
so(4).

The definition of the Euclidean version (\ref{VEV2}) via $\pi_{\sigma^2}\!$
$\rightarrow$ $\!-i\pi_{\sigma^2}$, as well as of Minkowski\-an one, contains
oscillating exponent. It is possible to reproduce the results of above considered
calculation of area monomial VEVs through $\delta$-functions of (antisymmetric part
of) the curvature by integrating monomials with monotonic exponent in terms of genuine
$\pi_{\sigma^2}$ by moving integration contour over curvature to complex plane
\cite{Kha5}. This contour should start at $\pmpi_{\sigma^2}\circ\pmR_{\sigma^2}\!$ =
$\!+|\pmbpi_{\sigma^2}|$, not at $\pmpi_{\sigma^2}\circ\pmR_{\sigma^2}\!$ =
$\!-|\pmbpi_{\sigma^2}|$. If $\pmR$ appears in the exponential in the form
$\pmpi\circ\pmR$, then appropriate complex change of variable $\pmbphi$ parameterizing
$\pmR$ corresponds to
\begin{eqnarray}%
\label{phi-i-eta}                                                                  
\pmphi & = & {\pi \over 2} + i\pmeta, ~~~ -\infty < \pmeta < +\infty, \\%
\label{theta-i-zeta}                                                               
\pmtheta & = & i\pmzeta, ~~~~~~~~~~~~~ 0 \leq \pmzeta < +\infty
\end{eqnarray}

\noindent where $\pmtheta$ is the azimuthal angle of $\pmbphi$ w.r.t. $\pmbpi$, the
polar angle $\pmchi$ remaining the same.

Now generalize (\ref{phi-i-eta}), (\ref{theta-i-zeta}) to the case when $\pmR$ enters
in the form $\pmm \circ \pmR$ where $\pmm$ has not only antisymmetric, but also scalar
part,
\begin{equation}                                                                   
\pmm = {1 \over 2}\pmbm \cdot \pmbSig + {1 \over 2}\pmm_0 \cdot 1.
\end{equation}

\noindent Of course, in this case $\pmm$ can not be (anti)selfdual part of anything
nor (anti)selfdual matrix itself. Here index $\pm$ means simply that it is sum of
products of (anti)selfdual matrices. The latter arise when we express curvatures on
$t$-like triangles in terms of independent ones with the help of Bianchi identities.
These curvatures can depend on the given $\Rsig^{\pm 1}$ linearly or not depend at
all. Therefore $\pmmsig \circ \pmRsig$ is the general form of dependence on the given
$\pmRsig$ in the exponential of (\ref{VEV2}). General form of integral over given
curvature matrix is
\begin{equation}                                                                   
\int \exp (-\pmbm\pmbn\sin \!\pmphi - \pmm_0 \cos \!\pmphi
){\sin^2 \!\pmphi \over \pmphi^2} \d^3 \pmbphi,
\end{equation}

\noindent where, remind, $\pmbn$ = $\pmbphi/\pmphi$, and azimuthal angle of $\pmbphi$
w.r.t. $\pmbm$ is $\pmtheta$. Apply (\ref{phi-i-eta}) and then ($\ref{theta-i-zeta}$)
to the {\it shifted} $\pmphi$,
\begin{eqnarray}                                                                   
& & \pmtheta = i\pmzeta, ~~~ \pmphi + \pmalpha = {\pi \over 2} +
i\pmeta, \\ & & \cos \!\pmalpha = {\sqrt{\pmbm^2} \cosh
\!\pmzeta\over \sqrt{\pmbm^2\cosh^2 \!\pmzeta + \pmm^2_0}},
~~~ \sin \!\pmalpha = {\pmm_0 \over \sqrt{\pmbm^2\cosh^2 \!\pmzeta + \pmm^2_0}}.   
\end{eqnarray}

\noindent The general case of complex $\pmbm$, $\pmm_0$ is implied. Important is that
$\pmSig_k$ are real-valued so that orthogonal conjugation operation is commuting with
analytic continuation. The branch of the function $\sqrt{z}$ is chosen in the complex
plane of $z$ with cut along negative real half-axis such that $\sqrt{1}$ = 1. (In
particular, this means that $\Re \sqrt{z}$ $\!\geq\!$ 0.) The integral over $\pmeta$,
$\pmzeta$ transforms to give
\begin{equation}\label{K1m}                                                        
\int \exp (-\pmm \circ \pmR) {\sin^2 \!\pmphi\over \pmphi^2} \d^3
\pmbphi = {4\pi \over \sqrt{{\rm tr}\pmbarm\pmm}}K_1\left
(\sqrt{{\rm tr}\pmbarm\pmm}\right ).
\end{equation}

\noindent The $K_1$ is the modified Bessel function.

The idea is to try to find some set of the 2-simplices ${\cal M}$ so that exponential
in (\ref{VEV2}) be representable in the form
\begin{equation}\label{-mR-piR}                                                    
- \sum_{\sigma^2 \in {\cal M}} \msig \circ \Rsig - \sum_{\sigma^2 \not\in {\cal M}}
\pisig \circ \Rsig
\end{equation}

\noindent where $\msig$ = $\pisig$ + (linear in $\{ \tausig \}$ terms). The notation
$\{ \dots \}$ means "the set of \dots ". The set $\{ \msig \}$ depend on $\{ \tausig
\}$ and on $\{ \Rsig | \sigma^2 \not\in {\cal M} \}$, but not on $\{ \Rsig | \sigma^2
\in {\cal M} \}$. Then integrations over $\{ \Rsig | \sigma^2 \in {\cal M} \}$ can be
explicitly performed according to eq. (\ref{K1m}) giving
\begin{eqnarray}\label{dN1}                                                        
\d \pmmu_{\rm area} & \equiv & \d \pmcalN \prod_{\stackrel{\stackrel{\rm not} {t-{\rm
like}}}{\sigma^2}}{\rm d}^3 \pmbpi_{\sigma^2}, \nonumber\\ & & \hspace{-28mm} \d
\pmcalN \Longrightarrow \left [ \prod_{\sigma^2 \in {\cal M}}{K_1\left (\sqrt{{\rm
tr}\pmbarmsig\pmmsig}\right ) \over \sqrt{{\rm tr}\pmbarmsig\pmmsig}} \right ] \exp
\left ( -\sum_{\stackrel{\stackrel{\rm not} {t-{\rm like}}}{\sigma^2 \not \in {\cal
M}}} |\pmbpisig |\cosh \!\pmzetasig \cosh \!\pmetasig \right ) \nonumber
\\ & & \hspace{-10mm} \cdot \prod_{\stackrel{\stackrel{\rm not} {t-{\rm
like}}}{\sigma^2 \not \in {\cal M}}} \cosh^2 \!\pmetasig \d \cosh
\!\pmetasig \d \cosh \!\pmzetasig \d \pmchisig
\end{eqnarray}

\noindent where $\{ \msig | \sigma^2 \in {\cal M} \}$ depend on $\{ \pmetasig,
\pmzetasig, \pmchisig | \sigma^2 \not\in {\cal M} \}$ through $\Rsig$ parameterized by
these,
\begin{eqnarray}                                                                   
\pmRsig & = & -i\sinh \!\pmetasig + \pmbSig \cdot \!\pmbnsig \cosh \!\pmetasig,
\nonumber\\ \pmbnsig & = & {\pmbpisig \over | \pmbpisig |} \cosh \!\pmzetasig +
i(\sinh \!\pmzetasig )(\pmbeonesig \cos \!\pmchisig + \pmbetwosig \sin \!\pmchisig)
\end{eqnarray}

\noindent where $\pmbeonesig$, $\pmbetwosig$ together with $\pmbpisig / | \pmbpisig |$
form orthonormal triple.

Rewrite (\ref{dN1}) as
\begin{eqnarray}\label{dN2}                                                        
\d \pmcalN & \Longrightarrow & \exp \left ( -\sum_{\sigma^2 \in {\cal M}} \sqrt{{\rm
tr}\pmbarmsig\pmmsig}\cosh \!\pmzetasig \cosh \!\pmetasig \right. \nonumber\\ & &
\hspace{-28mm} \left. -\sum_{\stackrel{\stackrel{\rm not} {t-{\rm like}}}{\sigma^2
\not \in {\cal M}}} |\pmbpisig |\cosh \!\pmzetasig \cosh \!\pmetasig \right )
\prod_{\stackrel{\stackrel{\rm not} {t-{\rm like}}}{\sigma^2}} \cosh^2 \!\pmetasig \d
\cosh \!\pmetasig \d \cosh \!\pmzetasig \d \pmchisig
\end{eqnarray}

\noindent where abstract dummy variables $\{ \pmetasig, \pmzetasig, \pmchisig |
\sigma^2 \in {\cal M} \}$ and integrations over them are introduced to represent $K_1$
differently from what is given by equation (\ref{K1m}) read from right to left.
Remarkable is that it looks as path integral measure with {\it positive} (real part
of) effective action whereas general relativity action remains unbounded from below
upon formal Wick rotation. The price is that exponential in (\ref{dN2}) has imaginary
part, and positivity of the Euclidean measure (upon integrating out curvature
matrices) does not follow automatically as in the case of the usual field theory with
bounded action since explicitly real form of (\ref{dN2}) reads
\begin{eqnarray}\label{dN2cos}                                                     
\d \pmcalN & \Longrightarrow & \exp \left ( -\sum_{\sigma^2 \in {\cal M}} \Re
\sqrt{{\rm tr}\pmbarmsig\pmmsig}\cosh \!\pmzetasig \cosh \!\pmetasig \right.
\nonumber\\ & & \hspace{-28mm} \left. -\sum_{\stackrel{\stackrel{\rm not} {t-{\rm
like}}}{\sigma^2 \not \in {\cal M}}} |\pmbpisig |\cosh \!\pmzetasig \cosh \!\pmetasig
\right ) \cos \left( \sum_{\sigma^2 \in {\cal M}} \Im \sqrt{{\rm
tr}\pmbarmsig\pmmsig}\cosh \!\pmzetasig \cosh \!\pmetasig \right) \nonumber\\ & &
\hspace{-28mm} \cdot \prod_{\stackrel{\stackrel{\rm not} {t-{\rm like}}}{\sigma^2}}
\cosh^2 \!\pmetasig \d \cosh \!\pmetasig \d \cosh \!\pmzetasig \d \pmchisig
\end{eqnarray}

\noindent that is nonconstant in sign due to cosine. Below we speculate that
positivity should be expected in the most part of (if not in the whole) range of
variation of area tensors $\pisig$ if $\tausig$ are sufficiently small.

To construct the set ${\cal M}$, note that due to the Bianchi identities dependence on
the matrix $\Rsig$ on the given leaf/diagonal triangle $\sigma^2$ in the exponential
of (\ref{VEV2}) comes from all the triangles constituting together with this
$\sigma^2$ a closed surface. This is surface of the $t$-like 3-prism, one base of
which is just the given $\sigma^2$, the lateral surface consists of $t$-like triangles
and goes to infinity. In practice, replace this infinity by some lowest (initial) leaf
where another base $\sigma^2_0$ is located the tensor of which $\pi_{\sigma^2_0}$ is
taken as boundary value. Consider a variety of such prisms with upper bases $\sigma^2$
placed in the uppest (final) leaf such that any link in this leaf belongs to one and
only one of these bases. That is, lateral surfaces of different prisms do not have
common triangles. Then the terms $\msig \circ \Rsig$ in (\ref{-mR-piR}) represent
contribution from these prisms, ${\cal M}$ being the set of their bases in the uppest
leaf.

To really reduce the measure to such form, we should express the curvature matrices on
the $t$-like triangles in terms of those on the leaf/diagonal ones. The curvature on a
leaf/diagonal triangle $\sigma^2$ as product of $\Omega$s includes the two matrices
$\Omega$ on the $t$-like tetrahedrons $\sigma^3$ adjacent to $\sigma^2$ from above and
from below. Knowing curvatures on the set of leaf/diagonal triangles inside any t-like
3-prism allows to successively express matrix $\Omega$ on any $t$-like tetrahedron
inside the prism in terms of matrix $\Omega$ on the uppest $t$-like tetrahedron in
this prism taken as boundary value. Expressions for the considered curvatures look
like (fig.\ref{3prism})
\begin{figure}\unitlength 0.20mm
\begin{picture}(200,200)(-150,20)
\put (150,20){\line(0,1){120}} \put (150,20){\line(3,2){120}} \put
(150,20){\line(4,1){160}} \put (150,60){\line(1,1){120}} \put
(150,60){\line(2,1){160}} \put (150,60){\line(3,1){120}} \put
(150,60){\line(1,0){160}} \put (150,140){\line(3,1){120}} \put
(150,140){\line(1,0){160}} \put (270,100){\line(0,1){80}} \put
(270,100){\line(1,-1){40}} \put (270,180){\line(1,-1){40}} \put
(270,180){\line(1,-3){40}} \put (310,60){\line(0,1){80}} \put (310,135){$~l^{+}$} \put
(270,180){$~k^{+}$} \put (137,135){$i^{+}$} \put (142,55){$i$} \put (270,100){$~k$}
\put (310,55){$~l$} \put (132,15){$i^{-}$}
\end{picture}
\renewcommand{\baselinestretch}{1.0}
\caption{Fragment of the $t$-like 3-prism.}\label{3prism}
\end{figure}
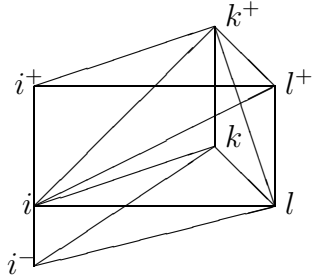
\begin{eqnarray}                                                                   
& & \hspace{-18mm} \dots \dots \dots \dots \dots \dots \dots \dots \dots \dots \dots
\nonumber\\ R_{(ikl)} & = & \dots \Om_{(i^-ikl)} \dots \Omega_{(ik^+kl)} \dots
\nonumber\\ R_{(ik^+)} & = & \dots \Om_{(ik^+kl)} \dots \Omega_{(ik^+l^+l)} \dots \\
R_{(ik^+l^+)} & = & \dots \Om_{(ik^+l^+l)} \dots \Omega_{(i^+ik^+l^+)} \dots
\nonumber\\ & & \hspace{-18mm} \dots \dots \dots \dots \dots \dots \dots \dots \dots
\dots \dots \nonumber
\end{eqnarray}

\noindent The dots in expressions for $R$ mean matrices $\Omega$ on the leaf/diagonal
tetrahedrons which can be considered as gauge ones. We can step-by-step express
$\Omega_{(i^-ikl)}$ $\rightarrow$ $\Omega_{(ik^+kl)}$ $\rightarrow$
$\Omega_{(ik^+l^+l)}$ $\rightarrow$ $\Omega_{(i^+ik^+l^+)}$ $\rightarrow$ \dots where
the arrow means "in terms of". Knowing $\Omega$s on $t$-like tetrahedrons we can find
the curvatures on $t$-like triangles, the products of these $\Omega$s,
\begin{equation}                                                                   
R_{(i^+ikl)} = \Omega^{\epsilon_{(ikl_n)l_{n-1}}}_{(i^+ikl_n)} \dots
\Omega^{\epsilon_{(ikl_1)l_n}}_{(i^+ikl_1)}.
\end{equation}

\noindent Here $\epsilon_{(ikl)m}$ = $\pm 1$ is some sign function. Thereby we find
contribution of the $t$-like triangles in terms of independent curvature matrices (on
the leaf/diagonal triangles).

In the continuum path integral formalism, one usually imposes boundary (initial/final)
conditions to unambiguously define the measure. Consideration of the two previous
paragraphs says that in our case fixing the initial leaf area tensors
$\pi_{\sigma^2_0}$ and final connections on the $t$-like tetrahedrons is appropriate.
Thereby, in particular, nontrivial integrations reduce to those over matching each
other sets $\d^6 \pisig$ and ${\cal D}^6 \Rsig$ on the leaf/diagonal $\sigma^2$.

Note an important particular case when integrations in (\ref{dN1}) are made over the
whole sets $\{ \pisig | \sigma^2 \not\in {\cal M} \}$ and $\{ \Rsig | \sigma^2 {\rm ~
is ~ not ~} t{\rm -like} \}$. The resulting measure factorizes over the 3-prisms with
upper bases constituting ${\cal M}$,
\begin{eqnarray}\label{factoriz}                                                   
\d \pmmu_{\rm area} \Longrightarrow \prod_{\sigma^2 \in {\cal M}}{K_1\left (\sqrt{{\rm
tr}\pmbarmsig\pmmsig}\right ) \over \sqrt{{\rm tr}\pmbarmsig\pmmsig}} \d^3
\pmbpi_{\sigma^2}.
\end{eqnarray}

\noindent Here $\msig$ is taken at $\{ \Rsig = 0 |  {\rm ~ not ~} t{\rm -like ~ }
\sigma^2 \not\in {\cal M} \}$ and differs from $\pisig$ by a constant, $\msig = \pisig
- \pisig^{(0)}$. In turn, $\pisig^{(0)}$ differs from the initial area tensor
$\pi_{\sigma^2_0}$ by the lateral 3-prism surface tensor: $\pisig^{(0)} -
\pi_{\sigma^2_0}$ is algebraic sum of tensors $\tau_{\sigma^2}$ for $\sigma^2$
constituting the lateral surface. The $\pisig^{(0)}$ has geometrical meaning of {\it
expected} value of area tensor $\pisig$ when the surface of the 3-prism closes due to
the (classical) equations of motion. The measure (\ref{factoriz}) describes quantum
fluctuation of $\pisig$ around $\pisig^{(0)}$. The (\ref{factoriz}) is explicitly
positive.

Thus, to represent exponential in (\ref{VEV2}) in the form (\ref{-mR-piR}) sufficient
is to divide the whole set of links in the uppest 3D leaf into triples forming the
triangles and take this set of triangles as ${\cal M}$ in (\ref{-mR-piR}). It is clear
that such set ${\cal M}$ does exist not for an arbitrary 3D leaf (at least the number
of links should be multiple of 3). In fig.\ref{M-simplex} probably the simplest
periodic cell of simplicial lattice is shown where the set ${\cal M}$ (shaded
triangles) is also periodic.

\newcounter{N}\unitlength 0.03mm
\begin{figure}
\begin{picture}(2000,1600)

\put(0,1000){\line(0,-1){500}}%

\put(0,1000){\line(2,1){800}}%
\put(0,500){\line(2,1){400}}%

\put(800,900){\line(0,-1){500}}%
\put(1000,1000){\line(0,-1){1000}}%

\put(500,500){\line(0,-1){500}}%

\put(0,1000){\line(1,0){500}}%

\put(500,0){\line(1,0){500}}%
\put(800,400){\line(1,0){1000}}%
\put(800,1400){\line(1,0){1000}}%
\put(800,900){\line(1,0){1000}}%

\put(900,1200){\line(1,0){500}}%
\put(500,1000){\line(2,1){800}}%
\put(1000,1000){\line(2,1){800}}%
\put(1400,700){\line(2,1){400}}%

\put(400,200){\line(2,1){400}}%
\put(400,200){\line(1,0){500}}%
\put(1000,0){\line(2,1){400}}%

\put(1300,1400){\line(0,-1){500}}%
\put(1400,1200){\line(0,-1){1000}}%
\put(1800,1400){\line(0,-1){1000}}%

\thicklines

\put(400,700){\line(1,1){500}}%
\put(400,700){\line(1,0){1000}}%
\put(800,1400){\line(1,-1){1000}}%
\put(800,1400){\line(1,-2){200}}%
\put(900,1200){\line(4,-3){400}}%
\put(900,1200){\line(0,-1){1000}}%
\put(400,1200){\line(1,0){500}}%
\put(400,1200){\line(0,-1){1000}}%
\put(400,700){\line(2,1){400}}%
\put(800,1400){\line(0,-1){500}}%
\put(500,1000){\line(0,-1){500}}%
\put(500,1000){\line(1,0){500}}%
\put(0,0){\line(1,1){1000}}%
\put(500,500){\line(2,1){800}}%
\put(0,0){\line(1,0){500}}%
\put(0,0){\line(2,1){400}}%
\put(900,200){\line(1,0){500}}%
\put(500,0){\line(2,1){800}}%
\put(1400,200){\line(2,1){400}}%
\put(0,500){\line(1,0){1000}}%
\put(1000,500){\line(2,1){400}}%
\put(400,700){\line(1,-2){100}}%
\put(1300,900){\line(1,-2){100}}%
\put(500,500){\line(4,-3){400}}%
\put(1000,1000){\line(4,-3){800}}%
\put(0,500){\line(0,-1){500}}%
\put(400,700){\line(1,-1){500}}%
\put(1300,900){\line(0,-1){500}}%
\put(900,200){\line(1,1){500}}%
\put(900,1200){\line(1,-1){500}}%

\thinlines

\put(820,1360){\line(1,-1){400}}%
\put(840,1320){\line(1,-1){300}}%
\put(860,1280){\line(1,-1){200}}%

\multiput(-5,0)(4,7){200}{.}%
\multiput(395,700)(9,2){100}{.}%
\put(-5,1000){.}%
\multiput(495,500)(4,7){100}{.}%

\multiput(495,500)(9,2){100}{.}%
\multiput(-5,0)(9,2){200}{.}%

\multiput(895,200)(4,7){100}{.}%

\setcounter{N}{400}

\multiput(850,700)(-50,0){7}{%
\addtocounter{N}{-40}%
\line(2,1){\value{N}}}

\setcounter{N}{400}

\multiput(800,945)(0,45){7}{%
\addtocounter{N}{-36}%
\line(-2,-1){\value{N}}}

\setcounter{N}{500}

\multiput(500,960)(0,-40){11}{%
\addtocounter{N}{-40}%
\line(1,0){\value{N}}}%

\setcounter{N}{500}

\multiput(860,680)(-40,-20){9}{%
\addtocounter{N}{-50}%
\line(0,1){\value{N}}}%

\setcounter{N}{510}

\multiput(433,1200)(43,0){11}{%
\addtocounter{N}{-43}%
\line(0,-1){\value{N}}}%

\put(920,1160){\line(1,-1){400}}%
\put(940,1120){\line(1,-1){300}}%
\put(960,1080){\line(1,-1){200}}%

\put(1320,860){\line(1,-1){400}}%
\put(1340,820){\line(1,-1){300}}%
\put(1360,780){\line(1,-1){200}}%

\put(420,660){\line(1,-1){400}}%
\put(440,620){\line(1,-1){300}}%
\put(460,580){\line(1,-1){200}}%

\setcounter{N}{400}

\multiput(950,500)(-50,0){7}{%
\addtocounter{N}{-40}%
\line(2,1){\value{N}}}

\setcounter{N}{400}

\multiput(1350,200)(-50,0){7}{%
\addtocounter{N}{-40}%
\line(2,1){\value{N}}}

\setcounter{N}{400}

\multiput(450,0)(-50,0){7}{%
\addtocounter{N}{-40}%
\line(2,1){\value{N}}}

\setcounter{N}{500}

\multiput(0,460)(0,-40){11}{%
\addtocounter{N}{-40}%
\line(1,0){\value{N}}}%

\setcounter{N}{500}

\multiput(900,660)(0,-40){11}{%
\addtocounter{N}{-40}%
\line(1,0){\value{N}}}%

\setcounter{N}{500}

\multiput(1260,380)(-40,-20){9}{%
\addtocounter{N}{-50}%
\line(0,1){\value{N}}}%

\setcounter{N}{500}

\multiput(360,180)(-40,-20){9}{%
\addtocounter{N}{-50}%
\line(0,1){\value{N}}}%

\end{picture}

\begin{picture}(2200,900)

\thicklines

\put(0,0){\line(1,1){500}}%
\put(0,0){\line(2,1){400}}%
\put(0,0){\line(1,0){500}}%
\put(0,0){\line(0,1){500}}%
\put(0,500){\line(1,0){500}}%
\put(500,0){\line(2,1){400}}%
\put(400,200){\line(0,1){500}}%
\put(400,700){\line(1,-2){100}}%
\put(400,700){\line(1,-1){500}}%
\put(500,500){\line(4,-3){400}}%

\thinlines

\put(0,500){\line(2,1){400}}%
\put(500,500){\line(2,1){400}}%
\put(400,200){\line(1,0){500}}%
\put(400,700){\line(1,0){500}}%
\put(500,0){\line(0,1){500}}%
\put(900,200){\line(0,1){500}}%

\put(420,660){\line(1,-1){400}}%
\put(440,620){\line(1,-1){300}}%
\put(460,580){\line(1,-1){200}}%

\setcounter{N}{500}

\multiput(360,180)(-40,-20){9}{%
\addtocounter{N}{-50}%
\line(0,1){\value{N}}}%

\setcounter{N}{500}

\multiput(0,460)(0,-40){11}{%
\addtocounter{N}{-40}%
\line(1,0){\value{N}}}%

\setcounter{N}{400}

\multiput(450,0)(-50,0){7}{%
\addtocounter{N}{-40}%
\line(2,1){\value{N}}}

\multiput(-5,0)(9,2){100}{.}%
\multiput(-5,0)(4,7){100}{.}%

\put(1200,0){\line(2,1){400}}%
\put(1200,500){\line(2,1){400}}%
\put(1700,0){\line(2,1){400}}%
\put(1700,500){\line(2,1){400}}%
\put(1200,0){\line(1,0){500}}%
\put(1200,500){\line(1,0){500}}%
\put(1600,200){\line(1,0){500}}%
\put(1600,700){\line(1,0){500}}%
\put(1200,0){\line(0,1){500}}%
\put(1700,0){\line(0,1){500}}%
\put(1600,200){\line(0,1){500}}%
\put(2100,200){\line(0,1){500}}%

\multiput(1195,500)(18,4){50}{.}%
\multiput(1695,0)(8,14){50}{.}%
\multiput(1695,0)(-7,14){14}{.}%
\multiput(1595,200)(-16,12){25}{.}%
\multiput(1195,500)(11.5,-11.5){44}{.}%
\multiput(1595,200)(11.5,11.5){44}{.}%

\put(1020,350){+}%

\put(400,-150){C0}%
\put(1600,-150){C1}%

\end{picture}

\bigskip
\renewcommand{\baselinestretch}{1.0}
\caption{Periodic cell of the simplicial manifold with triangles marked (shaded) in
such a way that any edge does belong to one and only one of marked triangles. It
consists of $2\times 2\times 2$ building blocks of the two types C0, C1 alternating in
all three directions.} \label{M-simplex}
\end{figure}
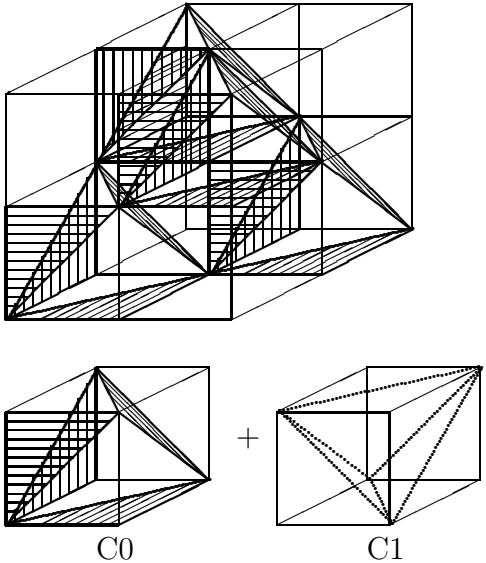
\unitlength 1pt%

Genuine simplicial decomposition possesses quite complex combinatorics, so let us
demonstrate main features of the result of above calculation by using as example the
cubic decomposition. The latter can be viewed as {\it sub}-minisuperspace of
simplicial system if one starts from the simplest periodic simplicial complex with
elementary 4-cubic cell divided by diagonals emitted from one of its vertices into 24
4-simplices \cite{RocWil}. Each 3-cube face built on three coordinate directions is
divided into 6 tetrahedrons, and we simply put $\Omega$s on these tetrahedrons to be
the same on the whole 3-cube. There are also the 3-cube faces built on two coordinate
and one diagonal direction, and we put $\Omega$s on the tetrahedrons forming these
faces to be 1. Each 2-face (square) is divided into two triangles, and the curvature
matrices on these triangles resulting from our choice of connections turn out to be
the same on this square and, besides, these differs from 1 only on the square built on
two coordinate directions, not on diagonal(s).

Introduce some cubic notations and definitions. By $\lambda$ we denote link in the
coordinate direction $\lambda$; $\lambda$, $\mu$, $\nu$, $\rho$, \dots = 1, 2, 3, 4.
Let the coordinate 4 be $t$. By $\sq$ denote a square. In particular, $\sq$ =
$|\lambda\mu|$ means the square built on the coordinate directions $\lambda$, $\mu$.
The connection matrix $\Omega_{\lambda}$ is that one on the 3-cube built on the
coordinates $\mu$, $\nu$, $\rho$ (and also denoted as $|\mu\nu\rho|$) complement to
$\lambda$. The set ${\cal M}$ for cubic decomposition of 3D leaf has periodic cell
consisting of 2 $\!\times\!$ 2 $\!\times\!$ 2 elementary cubes, see fig.\ref{cube}
corresponding to%
\unitlength 0.03mm
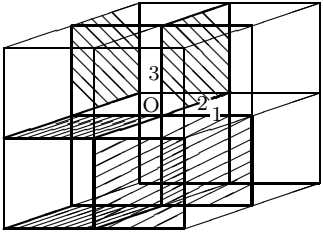
\begin{figure}
\begin{picture}(1600,1200)

\put(0,800){\line(0,-1){800}}%
\put(0,800){\line(3,1){600}}%
\put(0,800){\line(1,0){800}}%

\thicklines

\put(300,900){\line(0,-1){400}}%
\put(600,1000){\line(0,-1){400}}%
\put(0,400){\line(3,1){600}}%
\put(0,400){\line(1,0){800}}%
\put(300,500){\line(1,0){610}}%
\put(960,500){\line(1,0){140}}%
\put(400,400){\line(3,1){465}}%
\put(0,0){\line(3,1){300}}%
\put(0,0){\line(1,0){800}}%
\put(300,100){\line(1,0){800}}%
\put(400,0){\line(3,1){300}}%
\put(1000,1000){\line(0,-1){400}}%
\put(700,900){\line(0,-1){800}}%
\put(1100,500){\line(0,-1){400}}%
\put(400,400){\line(0,-1){400}}%
\put(700,900){\line(3,1){300}}%
\put(800,400){\line(0,-1){400}}%

\thinlines

\put(300,500){\line(0,-1){400}}%
\put(300,900){\line(1,0){800}}%
\put(600,600){\line(0,-1){400}}%
\put(600,1000){\line(1,0){800}}%
\put(400,800){\line(0,-1){400}}%
\put(400,800){\line(3,1){300}}%
\put(800,800){\line(0,-1){400}}%
\put(800,800){\line(3,1){600}}%
\put(600,600){\line(1,0){800}}%
\put(800,400){\line(3,1){600}}%
\put(300,100){\line(3,1){300}}%
\put(600,200){\line(1,0){800}}%
\put(700,100){\line(3,1){300}}%
\put(800,0){\line(3,1){600}}%
\put(1400,1000){\line(0,-1){800}}%
\put(1000,600){\line(0,-1){400}}%
\put(1100,900){\line(0,-1){400}}%

\setcounter{N}{100}

\multiput(900,500)(100,0){3}{%
\addtocounter{N}{100}%
\line(-2,-1){\value{N}}}

\multiput(1100,450)(0,-50){4}{%
\line(-2,-1){400}}%

\multiput(800,100)(100,0){2}{%
\addtocounter{N}{-100}%
\line(2,1){\value{N}}}

\setcounter{N}{50}

\multiput(550,400)(100,0){3}{%
\addtocounter{N}{100}%
\line(-2,-1){\value{N}}}%

\multiput(800,375)(0,-50){4}{%
\line(-2,-1){400}}%

\setcounter{N}{450}

\multiput(450,0)(100,0){3}{%
\addtocounter{N}{-100}%
\line(2,1){\value{N}}}%

\multiput(50,400)(50,0){7}{%
\line(3,1){300}}%

\multiput(50,0)(50,0){7}{%
\line(3,1){300}}%

\setcounter{N}{345}

\multiput(300,900)(0,-60){5}{%
\addtocounter{N}{-45}%
\line(1,-1){\value{N}}}

\setcounter{N}{300}

\multiput(600,660)(0,60){4}{%
\addtocounter{N}{-45}%
\line(-1,1){\value{N}}}

\setcounter{N}{345}

\multiput(700,900)(0,-60){5}{%
\addtocounter{N}{-45}%
\line(1,-1){\value{N}}}

\setcounter{N}{300}

\multiput(1000,660)(0,60){4}{%
\addtocounter{N}{-45}%
\line(-1,1){\value{N}}}

\put(855,525){\scriptsize 2}%
\multiput(905,570)(9,3){10}{.}%
\put(920,475){\scriptsize 1}%
\put(640,660){\scriptsize 3}%
\put(615,515){\scriptsize O}%

\end{picture}
\renewcommand{\baselinestretch}{1.0}
\caption{Periodic cell of the lattice with squares marked (shaded) in such a way that
any edge does belong to one and only one of marked squares. It contains $2\times
2\times 2$ elementary cells of the unmarked lattice.}\label{cube}

\end{figure}
\unitlength 1pt
\begin{equation}                                                                   
{\cal M} = \sum_{k_1, k_2, k_3} T^{2k_1}_1 T^{2k_2}_2 T^{2k_3}_3 \left (|23| + \T_1
|23| + \T_3 |31| + \T_{12} |12| + \T_{23} |31| + \T_{123} |12| \right )
\end{equation}

\noindent in the uppest leaf. Here $k_1$, $k_2$, $k_3$ are integers, $T_{\lambda}$ is
translation to the neighboring vertex in the direction $\lambda$. Expressions for the
measure follow from those for the simplicial case (\ref{dN1}) -- (\ref{dN2}) by
replacing $\sigma^2$ $\rightarrow$ $\sq$. There are several choices of the 4-cube
containing a given square in the frame of which tensor of this square is defined. If
the two area tensors are defined in the same frame, the result of integrating the
measure over connections will depend on the scalar constructed of these two tensors.
Therefore it seems to be a good idea to define area tensors in possibly different
frames, as in fig.\ref{frames}, case (b).%
\unitlength 0.04mm
\begin{figure}
\begin{picture}(1800,1200)

\put(568,580){$\odot$}%
\put(468,680){$\odot$}%
\put(668,680){$\odot$}%

\put(1168,580){$\odot$}%

\put(1268,680){$\odot$}%
\put(1068,680){$\odot$}%

\put(825,50){\vector(-1,2){310}}%
\put(975,50){\vector(1,2){310}}%

\put(840,0){(b)}%

\put(625,600){\line(1,0){550}}%
\put(600,575){\line(0,-1){550}}%
\put(1200,575){\line(0,-1){550}}%
\put(25,600){\line(1,0){550}}%
\put(1225,600){\line(1,0){550}}%
\put(600,635){\line(0,1){440}}%
\put(1200,635){\line(0,1){440}}%

\put(800,700){\vector(0,-1){200}}%
\put(1000,500){\vector(0,1){200}}%

\put(680,500){$\Omega_1$}%
\put(1020,490){$\Om_1$}%
\put(885,520){3}%

\put(860,1020){\vector(-1,-2){140}}%
\put(940,1020){\vector(1,-2){140}}%
\put(845,1050){(a)}%

\put(300,800){\vector(4,-1){365}}%
\put(300,800){\vector(2,-1){165}}%

\put(1500,800){\vector(-4,-1){365}}%
\put(1500,800){\vector(-2,-1){165}}%

\put(500,950){\vector(1,0){200}}%
\put(1100,950){\vector(1,0){200}}%
\put(750,350){\vector(-1,0){250}}%
\put(1300,350){\vector(-1,0){250}}%
\put(300,500){\vector(0,1){200}}%
\put(1500,700){\vector(0,-1){200}}%

\put(560,805){1}%
\put(1215,810){$T_31$}%
\put(150,800){$\tau_{|42|}$}%
\put(1515,800){$T_3\tau_{|42|}$}%
\put(490,975){$\Om_3$}%
\put(1220,975){$T_3\Om_3$}%

\setcounter{N}{700}

\multiput(600,700)(0,50){6}{%
\addtocounter{N}{-100}%
\line(2,1){\value{N}}}%

\put(1200,950){\line(-2,-1){470}}%
\put(1200,900){\line(-2,-1){570}}%
\put(1200,850){\line(-2,-1){500}}%
\put(1200,800){\line(-2,-1){400}}%
\put(1200,700){\line(-2,-1){200}}%
\put(1200,650){\line(-2,-1){100}}%
\put(900,600){\line(2,1){170}}%

\put(380,250){$\T_1\Omega_3$}%
\put(1220,250){$\T_1T_3\Omega_3$}%

\put(80,490){$\T_3\Om_1$}%
\put(1520,500){$T_3\Omega_1$}%

\put(500,500){2}%
\put(1250,500){$T_32$}%

\end{picture}
\renewcommand{\baselinestretch}{1.0}
\caption{To assigning to the squares $|42|$, $T_3|42|$ the frames of definition of
their area tensors $\tau_{|42|}$, $T_3\tau_{|42|}$. In the pictured current 3D leaf
these squares are observed as the links 2, $T_32$, respectively (shown perpendicular
to the plane of picture). The chosen frame is pointed out by slightly parallel moving
the given square to the chosen 4-cube (that is, parallel moving the given link to the
chosen 3-cube in the pictured 3D leaf). (a).$\tau_{|42|}$, $T_3\tau_{|42|}$ are
defined in the frame of the cube $|123|$ (shaded). (b). $\tau_{|42|}$,
$T_3\tau_{|42|}$ are defined outside the cube $|123|$.}\label{frames}

\end{figure}
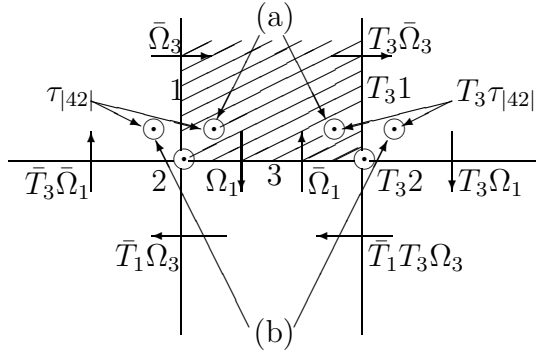
Of course, corresponding curvature matrices should be defined in the same frames. With
this rule of definition the curvature matrices on the lateral squares of, say, the
3-prism with base $|23|$ take the form $\T^n_4R_{\sq}$, $n$ = 1, 2, 3, \dots, $\sq$ =
$|42|$, $|43|$, $T_3|42|$, $T_2|43|$,
\begin{eqnarray}                                                                   
R_{|42|} & = & \left(\T_3\Om_1\right)\left(\T_1\Omega_3\right)\Omega_1\Om_3,
\nonumber\\ T_3R_{|42|} & = & \left( T_3\Om_3 \right) \Om_1 \left( T_3\T_1\Omega_3
\right) \left( T_3\Omega_1 \right), \nonumber\\ R_{|43|} & = & \Omega_2 \Om_1 \left(
\T_1 \Om_2 \right)\left( \T_2 \Omega_1 \right), \\ T_2R_{|43|} & = & \left( T_2 \Om_1
\right) \left( T_2\T_1\Om_2 \right) \Omega_1 \left( T_2\Omega_2 \right). \nonumber
\end{eqnarray}

\noindent (By default, notations $\Omega_{\lambda}$, $R_{|\lambda\mu|}$ are referred
to the uppest leaf.) Denote $\Omega_4$ $\equiv$ $U$. The matrices
$\T^n_4\Omega_{\alpha}$, $n$ = 1, 2, 3, \dots, $\alpha$, $\beta$, $\gamma$, \dots = 1,
2, 3 can be found in terms of $\Omega_{\alpha}$, $\T^k_4R_{|\alpha\beta|}$, $\T^k_4U$,
$k$ = 0, 1, \dots, $n-1$ from
\begin{eqnarray}                                                                   
R_{|23|} & = & \U \left( \T_4 \Om_1 \right) \left( \T_1 U \right) \Omega_1,
\nonumber\\ & & \hspace{-15mm} \dots {\rm cycle ~ perm ~ 1, 2, 3} \dots .
\end{eqnarray}

\noindent Thereby we find contribution of the $t$-like squares in terms of independent
curvature matrices (on the leaf squares) and eventually matrices $m_{\sq}$.

For $m_{|23|}$ the result reads
\begin{eqnarray}\label{m23}                                                        
& & \hspace{-20mm} m_{|23|} = \pi_{|23|} + \sum^N_{n=1} \left( \prod^{k=1}_{n-1}
\T^k_4 \R_{|23|} \right) \left\{ \left( \prod^{k=0}_{n-1} \T^k_4 T_3 \R_{|12|} \right)
\left( \T^n_4 T_3 \tau_{|42|} \right) \left( \prod^{n-1}_{k=0} \T^k_4 T_3 R_{|23|}
\right) \right. \nonumber \\ & & \hspace{-20mm} \cdot \left( \prod^{n-1}_{k=0} \T^k_4
T_3 \T_1 R_{|12|} \right)  - \left( \prod^{n-1}_{k=0} \T^k_4 R_{|12|} \right) \left(
\T^n_4 \tau_{|42|} \right) \left( \prod^{n-1}_{k=0} \T^k_4 \T_3 R_{|23|} \right)
\left( \prod^{k=0}_{n-1} \T^k_4 \T_1 \R_{|12|} \right) \nonumber \\ & & + \left(
\prod^{n-1}_{k=0} \T^k_4 R_{|31|} \right) \left( \T^n_4 \tau_{|43|} \right) \left(
\prod^{n-1}_{k=0} \T^k_4 \T_2 R_{|23|} \right) \left( \prod^{k=0}_{n-1} \T^k_4 \T_1
\R_{|31|} \right) \\ & & \left. - \left( \prod^{k=0}_{n-1} \T^k_4 T_2 \R_{|31|}
\right) \left( \T^n_4 T_2 \tau_{|43|} \right) \left( \prod^{n-1}_{k=0} \T^k_4 T_2
R_{|23|} \right) \left( \prod^{n-1}_{k=0} \T^k_4 T_2 \T_1 R_{|31|} \right) \right\}
\nonumber
\end{eqnarray}

\noindent where $N$ + 1 is the number of leaves, and the products of matrices are
ordered according to the rule
\begin{equation}                                                                   
\prod^n_{k=0} A_k = A_n A_{n-1} \dots A_0.
\end{equation}

\noindent For other $m_{\sq}$ we cyclically permute 1, 2, 3 and translate in the
directions 1, 2, 3. For simplicity, here we have put equal to 1 the boundary values
$\Omega_{\alpha}$ on the uppest leaf and to zero boundary values $\pi_{\sq_0}$ on the
lowest leaf. Besides that, gauge matrices $\T^n_4U$, $n$ = 0, 1, 2, \dots are set to
be 1. This corresponds to extending the local frame from any 4-cube to the whole
t-like 4-prism containing this 4-cube.

At the point $\{ \pmzeta_{\sq} = 0, \pmeta_{\sq} = 0 | \sq \not\in {\cal M} \}$ where
the factor in the measure corresponding to contribution from the squares $\sq$
$\not\in$ ${\cal M}$ reaches its maximum, and for uniform orthogonal lattice take
$\pmpi_{|23|}$ = $A \pmSig_1/4$, $\pmtau_{|41|}$ = $\pm \varepsilon \pmSig_1/4$, \dots
cycle permutations of 1, 2, 3 \dots, then $\pmR_{|23|}$ = $\pmSig_1$, \dots. In
(\ref{m23}) we find sum of sign-altered terms so that $m_{|23|} = \pi_{|23|} +
O(\varepsilon)$. For estimate, let $A$, $\varepsilon$ be typical areas of the leaf and
$t$-like squares, respectively.
In the explicitly real expression for the measure (\ref{dN2cos}) to be integrated, the
cosine may become negative if for some variable $\zeta$ or $\eta$ we have $\sinh \zeta
= O(A/\varepsilon)$ or $\sinh \eta = O(A/\varepsilon)$. However, contribution from
negative half-wave of cosine to the entire integral over $\zeta, \eta$-variables is
dumped by the factor $\exp (-O(A^2/\varepsilon))$ in this case. Therefore at $A \geq
A_0 = O(\sqrt{\varepsilon})$ contribution of the negative half-waves of cosine is
dominated by positive ones, and resulting $\pmcalN$ is positive.

This is quite rough, sufficient estimate. In reality, the region of positivity of
$\pmcalN$ well may be larger then this or even coincide with the whole range of
varying the area tensors. The one-dimensional example is inequality $\int^{\infty}_0
f(x) \cos x \d x > 0$ which can be easily proved to hold for {\it any} concave
function ($f''(x)\!$ $\!>\!$ 0; in particular, for $f(x)\!$ = $\!\exp (-kx)$ at {\it
any} $k > 0$). And even the inequality $\pmcalN\!$ $\!>\!$ 0 is, generally speaking,
redundant for positivity means only $\pcalN \mcalN$ $>$ 0. Especially this
circumstance is expected to promote the measure be positive when $\pcalN$ and $\mcalN$
are dependent. This takes place on the physical hypersurface singled out by the
relations on area tensors of the type (\ref{v*v}) which connect $\pvsig$ and $\mvsig$.

Thus, completely discrete version of path integral in simplicial gravity can be
naturally formulated with some boundary (initial/final) conditions. Representation of
simplicial general relativity action in terms of area tensors and finite rotation
matrices (connection and curvature) is used. Discrete connection and curvature on
classical solutions of the equations of motion are not, strictly speaking, genuine
connection and curvature, but more general quantities and, therefore, these do not
appear as arguments of a function to be averaged, but are the integration (dummy)
variables. Despite of unboundedness of general relativity action, path integral can be
written in the form resembling that with positive (real part of) effective action by
moving integration contours over curvature to complex plane. This effective action is
not purely real, but arguments are given that the resulting path integral measure is
expected to be positively defined upon integrating over connection matrices. Up to
some integrable factor, this measure is dominated by the product of exponentially (in
area) falling off factors on separate areas.

It is interesting that our arguments use simplicial structure although built in a
simple regular way of similar 3-dimensional leaves, but with rather complex structure
of these leaves themselves; simplest leaf will not do. The work to extend the results
to arbitrary structure is in order.

The present work was supported in part by the Russian Foundation for Basic Research
through Grant No. 05-02-16627-a.


\begin{thebibliography}{99}
\bibitem{Regge}
 T. Regge, General relativity theory without coordinates. - Nuovo Cimento {\bf 19}, 568 (1961).
\bibitem{Fried}
 R. Friedberg, T. D. Lee, Derivation of Regge's action from Einstein's theory of general
 relativity. - Nucl. Phys. B {\bf 242}, 145 (1984).
\bibitem{Fein}
 G. Feinberg, R. Friedberg, T. D. Lee, M. C. Ren, Lattice gravity near the continuum
 limit. - Nucl. Phys. B {\bf 245}, 343 (1984).
\bibitem{RegWil}
 T. Regge and R.M. Williams, Discrete structures in gravity. - Journ. Math. Phys., {\bf 41}, 3964 (2000), gr-qc/0012035.
\bibitem{HamWil1}
 H. Hamber and R.M. Williams, Newtonian Potential in Quantum Regge Gravity. - Nucl.Phys. B, {\bf 435}, 361 (1995), hep-th/9406163.
\bibitem{HamWil2}
 H. Hamber and R.M. Williams, On the Measure in Simplicial Gravity. - Phys. Rev. D, {\bf 59},  064014 (1999), hep-th/9708019.
\bibitem{Fro}
 J. Fr\"{o}hlich, Regge Calculus and Discretized Gravitational Functional
 Integrals, I.~H.~E.~S.~preprint 1981 (unpublished);  Non-Perturbative Quantum Field
 Theory: Mathematical Aspects and Applications, Selected Papers (World Scientific,
 Singapore, 1992), pp. 523-545.
\bibitem{Kha1}
 V. M. Khatsymovsky, Tetrad and self-dual formulations of Regge calculus. - Class. Quantum Grav. {\bf 6}, L249 (1989).
\bibitem{Kha2}
 V. M. Khatsymovsky, Regge calculus in the canonical form. - Gen. Rel. Grav. {\bf 27}, 583 (1995), gr-qc/9310004.
\bibitem{Kha3}
 V. M. Khatsymovsky, A version of quantum measure in Regge calculus in three dimensions.
 - Class. Quantum Grav. {\bf 11}, 2443 (1994), gr-qc/9310040.
\bibitem{BarRocWil}
 J.W. Barrett, M. Ro\v{c}ek and R.M. Williams, A note on area variables in Regge calculus.
 - Class. Quantum Grav., {\bf 16}, 1373
 (1999), gr-qc/9710056.
\bibitem{Kha4}
 V. M. Khatsymovsky, Area expectation values in quantum area Regge cal\-culus. - Phys. Lett. {\bf 560B}, 245 (2003), gr-qc/0212110.
\bibitem{Kha7}
 V. M. Khatsymovsky, Length expectation values in quantum Regge calculus. - Phys. Lett. {\bf 586B}, 411 (2004), gr-qc/0401053.
\bibitem{Wael}
 H. Waelbroeck, 2+1 lattice gravity. - Class. Quantum Grav. {\bf 7}, 751 (1990).
\bibitem{Kha5}
 V. M. Khatsymovsky, Path integral in area tensor Regge calculus and complex
 connections. - Phys. Lett. {\bf 637B}, 350 (2006), gr-qc/0602116.
\bibitem{RocWil}
 M. Rocek and R.M. Williams, Quantum Regge calculus. - Phys. Lett. {\bf 104B}, 31
 (1981).

\end{thebibliography}
\end{document}